\documentclass[aps,prl,twocolumn,showpacs,longbibliography]{revtex4-2}
\usepackage{tikz} 
\usetikzlibrary{shapes,arrows,positioning,automata,backgrounds,calc,er,patterns}
\usepackage{graphicx}
\usepackage{dcolumn}
\usepackage{bm}
\usepackage{amsmath}
\usepackage{amssymb}
\usepackage{amsthm}
\usepackage{amsfonts}
\usepackage{enumerate}
\usepackage{siunitx}
\usepackage[colorlinks=true,urlcolor=blue,citecolor=blue,linkcolor=blue]{hyperref}

\definecolor{purple}{rgb}{0.8,0,0.6}
\definecolor{darkgreen}{rgb}{0.00,0.6,0.00}

\begin{document}

\title{Anomalous chiral magnetic effect in time reversal symmetry breaking Weyl semimetals}

\author{Long Liang}
\affiliation{Nordita, KTH Royal Institute of Technology and Stockholm University, Roslagstullsbacken 23, SE-106 91 Stockholm, Sweden}

\begin{abstract}
    We propose a mechanism to generate dissipationless current in time reversal symmetry breaking Weyl semimetals through the anomalous chiral magnetic effect (ACME). 
    The ACME current is induced by chiral imbalance and flows along the direction of the Weyl nodes separation in momentum space. 
    In contrast to the chiral magnetic effect, the ACME  is not related to the chiral anomaly and does not require external magnetic field. 
    In the presence of parallel electric and magnetic fields, the ACME gives rise to an antisymmetric linear magnetoresistance and a planar Hall conductivity, which we estimate to be observable.
\end{abstract}

\maketitle

\emph{Introduction.\textemdash}Topological semimetals have emerged as a new research frontier during the past couple of years~\cite{Franz:book-2013,Vafek-Vishwanath:rev-2014,Wehling-Balatsky:rev-2014,Armitage:rev-2018,HongDingRMP,yu2021pseudogauge}. 
The best known examples are Dirac and Weyl semimetals, whose characteristic features are the gapless nodes and linear dispersion relation around the nodes. 
They provide a versatile playground to investigate the interplay between relativistic particles and exotic emergent fields.

Current transport is one of the fundamental phenomena in physics. 
Since the current changes sign under time reversal ($\mathcal{T}$) or inversion ($\mathcal{P}$) transformation, to generate an electric current, both $\mathcal{T}$ and $\mathcal{P}$ symmetries must be broken. 
For example, in the Ohm's law, the electric field breaks $\mathcal{P}$, and the conductivity, related to dissipation processes in materials, breaks $\mathcal{T}$. 
The Hall effect provides an example of dissipationless current, and in this case, $\mathcal{T}$ is broken by the applied magnetic field.

The nontrivial topology and nodal degrees of freedom in Weyl semimetals allow for new possibilities of current generation. 
One mechanism that attracts great research interest from both condensed matter and high energy communities is the chiral magnetic effect (CME)~\cite{PhysRevD.22.3080,PhysRevD.78.074033,PhysRevB.85.165110,PhysRevLett.109.181602,Zhou_2013,Kharzeev_2014}, where a dissipationless  current is generated by $\mathcal{T}$-breaking magnetic field $\mathbf{B}$ and $\mathcal{P}$-breaking chiral chemical potential $\mu_5$, $\mathbf{j}\propto \mu_5\mathbf{B}$.
This effect was first predicted by using  field theoretical arguments related to the chiral anomaly~\cite{PhysRev.177.2426,Bell1969}. 
Based on the  CME, the negative magnetoresistance has been predicted~\cite{NIELSEN1983389,PhysRevB.88.104412} and observed~\cite{2015Sci...350..413X,Li_2015,Li_2016,PhysRevX.5.031023,Arnold_2016}.

In this Letter, we reveal a  mechanism to generate dissipationless electric current in $\mathcal{T}$-broken Weyl semimetals.
The current is proportional to $\mu_{5}$ and flows along the Weyl nodes separation in the momentum space $2\mathbf{k}_W$, 
\begin{eqnarray}
    \mathbf{j}(\omega) \sim \mu_{5}(\omega)\mathbf{k}_W.
\end{eqnarray}
Since the external magnetic field is not required, we call this phenomenon the \emph{anomalous chiral magnetic effect} (ACME).
The current survives in the zero frequency or dc limit, and depends weakly on $\omega$ in the low frequency limit.
In the ACME, $\mathcal{T}$ is broken intrinsically by the Weyl nodes separation instead of irreversible processes in materials, thereby making the effect free of dissipation in nature~\cite{Kharzeev_2014}.
We emphasize that the ACME stems from the symmetry rather than the topology of the system, and the current is nonvanishing even if the Weyl points are gapped out. 
In this sense the effect is general and could be realized in other materials than Weyl semimetals.

\begin{figure}[t]
	\includegraphics[width=0.35\textwidth]{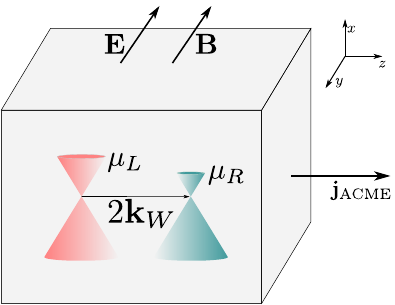}
	\caption{Schematic experimental setup for the ACME. 
		The chiral imbalance can be manipulated through various techniques. 
		As an example, here we use parallel electric and magnetic fields to induce a nonzero chiral chemical potential. 
		The ACME current flows along the direction of Weyl nodes separation and leads to a linear magnetoresistance or a planar Hall conductivity, depending on the direction of the applied fields.
	}\label{Fig:setup}
\end{figure}

The ACME could be realized in various setups where the chiral imbalance can be manipulated. 
For example, a chiral chemical potential can be induced by chiral anomaly~\cite{PhysRevB.88.104412}, phonon dynamics~\cite{PhysRevB.94.214306,Yuan:2020nwq}, strain~\cite{PhysRevB.94.241405}, and light~\cite{PhysRevB.102.245123,Cheng_2021}.
Recently, magnetic Weyl semimetals 
have been discovered~\cite{Belopolski1278,Liu1282,Morali1286,Wang_2019,Soh_2019,Ma_2019,doi:10.1063/1.5129467,PhysRevB.101.140402}, providing opportunities to verify our prediction.
Figure~\ref{Fig:setup} shows a possible experimental setup to detect the effect. 
Parallel electric and magnetic fields are applied to a $\mathcal{T}$--broken Weyl semimetal.
A chiral chemical potential is generated through the chiral anomaly, and a current along the direction of the Weyl nodes separation will be induced. 
Depending on whether $\mathbf{E}$ and $\mathbf{B}$ are parallel  to the Weyl nodes separation or not, the ACME current results in a linear magnetoresistance (LMR) such that the resistance becomes antisymmetric with respect to the magnetic field~\cite{PhysRevB.94.241105,PhysRevB.95.245128,PhysRevB.96.045112,PhysRevB.101.201410,zyuzin2021linear,Wang_2020,PhysRevLett.126.236601}, or a planar Hall conductivity~\cite{PhysRevB.96.041110,PhysRevLett.119.176804,PhysRevB.97.201110,PhysRevB.98.161110,PhysRevB.99.115121,Wang_2020,PhysRevLett.126.236601} that is nonzero for colinear electric and magnetic fields. Our estimate shows that the effect is large enough to be observed.

\emph{Model.\textemdash}We start from the widely used four band model defined through the momentum space Hamiltonian~\cite{PhysRevB.84.235126,PhysRevB.86.115133,PhysRevB.93.085442,PhysRevD.86.045001}
\begin{equation}\label{Eq:Hamiltonian}
    H(\mathbf{k})=\sum_{i}(k_i+b_i\gamma_5)\gamma^0\gamma^i+M\gamma^0,
\end{equation}
where $\gamma^\mu$ with $\mu=0,1,2,3$ are the Dirac gamma matrices satisfying anticommutation relations $\{\gamma^\mu,\gamma^\nu\}=2g^{\mu\nu}$ with $g^{\mu\nu}=\mathrm{diag}(1,-1,1,-1)$, $\gamma_5=i\gamma^0\gamma^1\gamma^2\gamma^3$ is the fifth gamma matrix, $\mathbf{k}$ is the momentum, $M$ is the mass parameter that is taken to be positive, and $\mathbf{b}$ is a Zeeman-like field~\cite{PhysRevB.84.235126} that breaks $\mathcal{T}$. 
The Fermi velocity $v_F$ and $\hbar$ are taken to be unity for simplicity and will be recovered when necessary. 
The Hamiltonian is invariant under $\mathcal{P}$ and particle-hole ($\mathcal{C}$) transformations, see the Supplemental Material (SM)~\cite{SI} for details.

\begin{figure}[t]
	\includegraphics[width=0.35\textwidth]{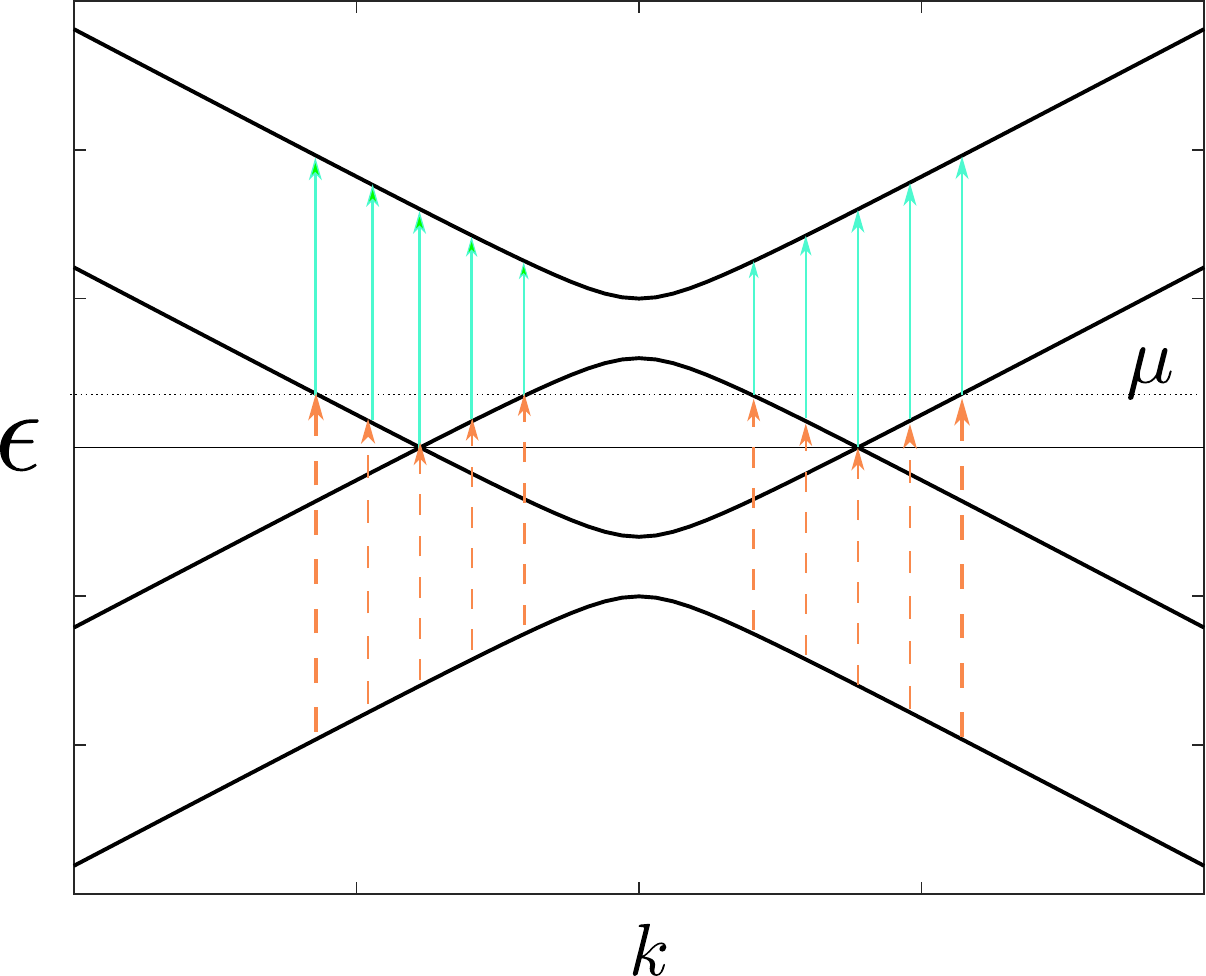}
	\caption{Depiction of the energy dispersion and
	interband transitions that contribute to the response function.
	}\label{Fig:interband_transitions}
\end{figure}

The  dispersion relations are given by
 $\epsilon_{s,\mathbf{k}}^2=\mathbf{k}^2+\mathbf{b}^2+M^2+2sA_\mathbf{k}$ with $s=\pm 1$ and $A_\mathbf{k}=\sqrt{(\mathbf{k}\cdot\mathbf{b})^2+M^2\mathbf{b}^2}$. %, and $B_{\mathbf{k}}=\mathbf{k}^2+\mathbf{b}^2+M^2$. 
The $s=1$ branch is gapped with the gap being %$\Delta_+=
 $M+|\mathbf{b}|$. For $|\mathbf{b}|>M$, the $s=-1$ branch becomes gapless at two Weyl points $\pm\mathbf{k}{_W}=\pm \sqrt{1-M^2/\mathbf{b}^2}\mathbf{b}$, and the dispersion is linear around these two points, see Fig.~\ref{Fig:interband_transitions}. With the decreasing of $|\mathbf{b}|$, the Weyl points are getting closer and finally annihilate when $|\mathbf{b}|=M$. Further decreasing $|\mathbf{b}|$, the energy spectrum becomes fully gapped. 

To induce an electric current, it is necessary to break $\mathcal{P}$. 
Physically, there are different $\mathcal{P}$--breaking terms~\cite{PhysRevB.84.235126} that can be added to the Hamiltonian, Eq.~\eqref{Eq:Hamiltonian}. 
Here we use the chiral chemical potential that can be manipulated in a variety of  setups~\cite{PhysRevB.88.104412,PhysRevB.94.214306,Yuan:2020nwq,PhysRevB.94.241405,PhysRevB.102.245123,Cheng_2021}, 
\begin{equation}\label{Eq:Hperturbation}
    H'=\mu_{5}(t,\mathbf{r})\gamma_5.
\end{equation}
Under the inversion transformation, $\gamma_5$ changes to $-\gamma_5$ and  $\mu_{5}(t,\mathbf{r})$  to $\mu_5(t,-\mathbf{r})$.
Therefore $\mathcal{P}$ is broken unless the chiral chemical potential is an odd function of position. 

A chiral chemical potential can also be introduced to describe the energy difference between the two Weyl nodes~\cite{PhysRevB.85.165110}. 
In this case, $\mu_5$ characterize an equilibrium property of the material, and there is no current in the ground state due to the generalized Bloch theorem~\cite{PhysRevD.92.085011}. 
To have a nonzero current, the chiral chemical potential should be treated as a perturbation such that the system is out of equilibrium~\cite{PhysRevB.88.125105,PhysRevLett.113.247203,PhysRevB.91.115203,Landsteiner_2016}.

Using the linear response theory, the induced current is related to the retarded current--chiral density correlation function $\Pi^{VA}_{i0}(\omega,\mathbf{q})$ through
\begin{eqnarray}
    j_i(\omega,\mathbf{q})=\Pi^{VA}_{i0}(\omega,\mathbf{q})\mu_{5}(\omega,\mathbf{q}).
\end{eqnarray}
We calculate the correlation function by using the imaginary time formalism,
\begin{eqnarray}
    \Pi^{VA}_{i0}(i\nu_m,\mathbf{q})=\frac{1}{\beta}\sum_{k}\mathrm{tr}\hat{j}_i(\mathbf{k}-\mathbf{q}/2) G(k) \gamma_5 G(k-q),~
\end{eqnarray}
where $\beta$ is the inverse temperature (the Boltzmann constant is taken to be unity), $k\equiv (i\omega_n,\mathbf{k})$, $\omega_n=(2n+1)\pi/\beta$ is the fermionic Matsubara frequency, and $\nu_m=2m\pi/\beta$ is the bosonic one.
The current operator is given by $\hat{j}_i(\mathbf{k})=\partial_{k_i}H(\mathbf{k})$, where the electric charge $e$ is set to be unity. 
The Matsubara Green's function reads
\begin{eqnarray}
    G(k)=\sum_{s,t=\pm 1}\frac{P_{s,t}(\mathbf{k})}{i\omega_n-t\epsilon_{s,\mathbf{k}}-\mu},
\end{eqnarray}
where $P_{s,t}(\mathbf{k})=|\psi_{s,t}(\mathbf{k})\rangle\langle \psi_{s,t}(\mathbf{k})|$ is the projection operator, $|\psi_{s,t}(\mathbf{k})\rangle$ is the eigenstate satisfying $H(\mathbf{k})|\psi_{s,t}(\mathbf{k})\rangle=t\epsilon_{s,\mathbf{k}}|\psi_{s,t}(\mathbf{k})\rangle$ with $t=\pm 1$, and $\mu$ is the chemical potential. 
A nonvanishing chemical potential breaks $\mathcal{C}$.
See the SM~\cite{SI} for the explicit expression of the Green's function.

It is worth pointing out that, although our theory is developed for a specific model, it is straightforward to generalize the calculation to other systems that break $\mathcal{T}$ and subjected to $\mathcal{P}$--violating perturbations. 
We thus expect that this current generation mechanism is general. 

\begin{figure*}[t]
	\includegraphics[width=0.8\textwidth]{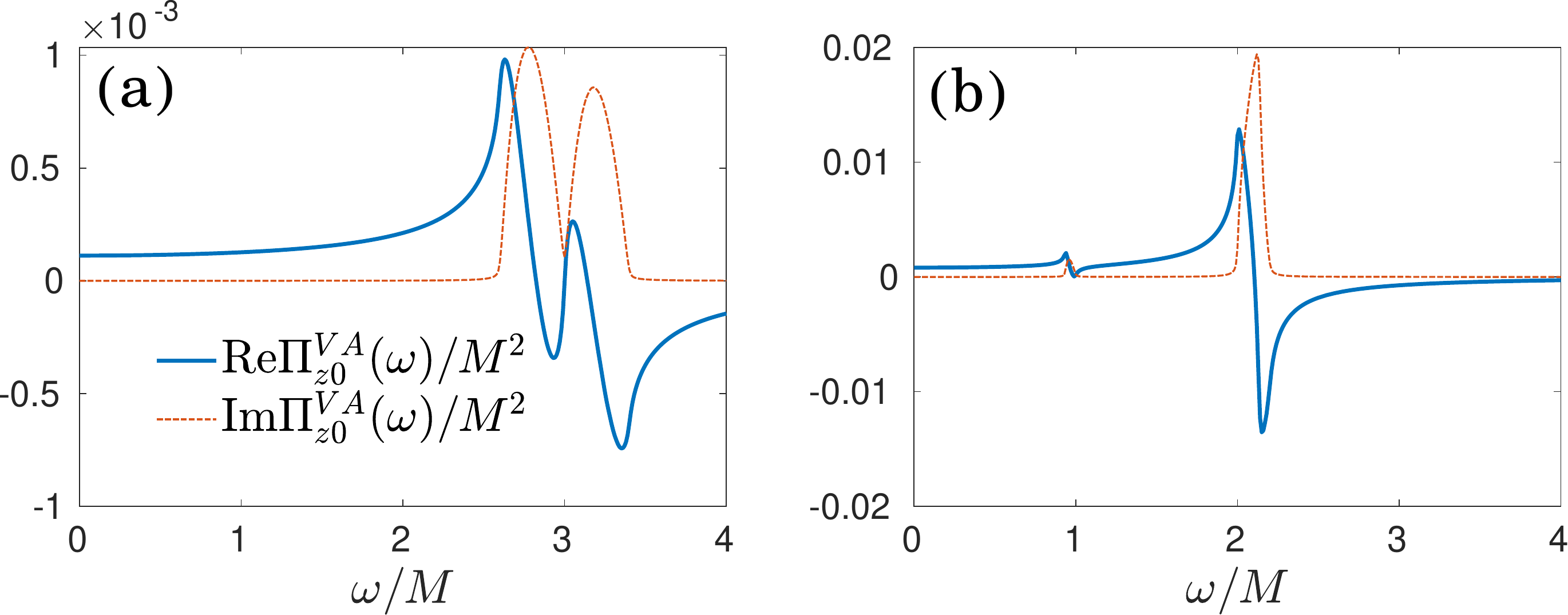}
	\caption{The frequency dependence of the response function $\Pi^{VA}_{z0}(\omega)$ in the Weyl semimetal (a) and gapped (b) phases. The time reversal symmetry breaking field $\mathbf{b}$ is along the $z$-direction. The parameters are $b_z/M=1.5$ and $\mu/M=0.2$ for (a) and $b_z/M=0.5$ and $\mu/M=0.6$ for (b). }\label{Fig:Pii0_omega}
\end{figure*}

\emph{Anomalous chiral magnetic effect.\textemdash}We are interested in the transport property, so we calculate the response function in the uniform limit by setting $\mathbf{q}=0$~\cite{PhysRevB.88.125105,PhysRevB.91.115203}. 
Note that the current operator changes sign under $\mathcal{C}$ transformation, while the chiral density remains invariant~\cite{Peskin:1995ev}. 
Consequently, if the chemical potential vanishes such that  $\mathcal{C}$ is preserved in the problem, the response function vanishes identically.
In the following we assume that the chemical potential is positive. We find that~\cite{SI}
\begin{widetext}
\begin{eqnarray}\label{Eq:Pii0_VA_omega}
    \Pi^{VA}_{i0}(\omega)
    &=&	\sum_{\mathbf{k},s=\pm}
    \frac{(\epsilon_{-,\mathbf{k}}-s\epsilon_{+,\mathbf{k}})[f(\epsilon_{-,\mathbf{k}})-f(\epsilon_{+,\mathbf{k}})]}{(\omega+i0^+)^2-(s\epsilon_{+,\mathbf{k}} - \epsilon_{-,\mathbf{k}})^2}
    \bigg[
    \frac{\mathbf{b}^2-A_\mathbf{k}}{\epsilon_{-,{\mathbf{k}}}}
    +s\frac{\mathbf{b}^2+A_\mathbf{k}}{\epsilon_{+,{\mathbf{k}}}}
    \bigg]
    \frac{M^2b_i}{A^2_\mathbf{k}}.
\end{eqnarray}
\end{widetext}

Since the momentum transfer $\mathbf{q}$ is set to be zero, only the interband processes contribute to $\Pi^{VA}_{i0}(\omega)$.
Because of the $\mathcal{P}$ and $\mathcal{C}$ symmetries of the Hamiltonian, Eq.~\eqref{Eq:Hamiltonian}, transitions between states with opposite energies are forbidden, see the SM~\cite{SI}. 
To better understand the physical meaning of Eq.~\eqref{Eq:Pii0_VA_omega}, we assume that the chemical potential lies in the $\epsilon_-$ band, see Fig.~\ref{Fig:interband_transitions}. 
Then the the occupied states in the $\epsilon_-$ band can be excited to the $\epsilon_+$ band, as depicted by the solid vertical lines in Fig.~\ref{Fig:interband_transitions}. 
This type of transitions is captured by the $s=1$ term in Eq.~\eqref{Eq:Pii0_VA_omega}. 
Besides those processes, the negative energy bands can also be excited to the unoccupied states. 
Using the fact that the transitions from negative energy states to positive energy ones sum to zero for vanishing chemical potential, we can rewrite the transitions from negative energy bands to the unoccupied states in terms of the transitions from $-\epsilon_+$ band to the occupied states in $\epsilon_-$ band, see the dashed vertical lines in Fig.~\ref{Fig:interband_transitions}. 
These processes correspond to the $s=-1$ term in Eq.~\eqref{Eq:Pii0_VA_omega}.

In Fig.~\ref{Fig:Pii0_omega}, we show the numerical results of the correlation function, Eq.~\eqref{Eq:Pii0_VA_omega}, in both the Weyl semimetal and gapped phases. 
We assume $\mathbf{b}$ is along the $z$-direction. 
As can be seen from the figure,  the correlation functions in both phases are nonzero, indicating that the effect is determined by the symmetry rather than the topology of the system. 
The imaginary part of the correlation function is nonvanishing only in a finite frequency region in which the energy conservation condition, $\omega=\pm(s\epsilon_+-\epsilon_-)$, is satisfied. 
In contrast, the real part is nonzero even in the zero frequency limit, and depends weakly on $\omega$ for frequencies do not fulfill the energy conservation condition.

We have used the unbounded continuum model to derive the above result. 
One may wonder whether the effect disappears or not if a lattice model is used~\cite{PhysRevLett.111.027201}. 
To see why the lattice model does not change the result qualitatively, it is enough to examine the imaginary part of the response function. The real part can be obtained through the Kramers--Kronig relation. 
As mentioned before, only the states satisfying the energy conservation condition contribute to $\mathrm{Im}\Pi_{i0}(\omega)$, see Eq.~\eqref{Eq:Pii0_VA_omega} and Fig.~\ref{Fig:interband_transitions}; consequently, the response function is not affected by the unbounded high energy states in the continuum approximation. 
To further confirm this, we calculate the response function directly by using a lattice mode~\cite{PhysRevLett.111.027201,PhysRevLett.115.177202}, and find that the result takes a similar form as Eq.~\eqref{Eq:Pii0_VA_omega}, see the SM~\cite{SI} for details.

\emph{ACME in the dc limit.\textemdash}In the dc limit, the correlation function, Eq.~\eqref{Eq:Pii0_VA_omega}, reduces to~\cite{SI}
\begin{eqnarray}\label{Eq:Pi_i0_dc}
    \Pi^{VA}_{i0,\mathrm{dc}}
    &=&\sum_{\mathbf{k},s}f(\epsilon_{s,\mathbf{k}})\partial_{k_i}\rho_{5,s}(\mathbf{k}),
\end{eqnarray}
where $\rho_{5,s}(\mathbf{k})=\langle\psi_{s,+}(\mathbf{k}) |\gamma_5 |\psi_{s,+}(\mathbf{k}) \rangle$ is the momentum resolved chiral density corresponding to the state $|\psi_{s,+}\rangle$.  
We note that Eq.~\eqref{Eq:Pi_i0_dc} applies to both continuum and lattice models. 
This expression has a simple physical interpretation: 
The perturbation $H'$ modifies the energy dispersion by $\delta \epsilon_{s,\mathbf{k}}=\rho_{5,s}(\mathbf{k})\mu_5$, which in turn leads to a correction to the group velocity $\delta \mathbf{v}_{s,\mathbf{k}}=\partial_{\mathbf{k}}\delta \epsilon_{s,\mathbf{k}}$ that contributes to the electric current.

Integrating by parts, Eq.~\eqref{Eq:Pi_i0_dc} can be rewritten as a Fermi surface integral of $\rho_{5,s}(\mathbf{k})$. 
Therefore the ACME can be regarded as a Fermi surface property, which is essentially different from the CME, whose existence does not rely on the Fermi surface.

For the model we use, the chiral symmetry is explicitly broken by the mass term. 
As a result, the chiral density $\rho_{5,s}(\mathbf{k})$ depends on the momentum and the dc response function is thus nonzero.  
We note that in realistic Weyl semimetal  materials, the chiral symmetry is always broken, thereby ensuring the ACME a universal phenomenon in $\mathcal{T}$--broken  Weyl materials.

Since time reversal symmetry breaking is characterized by the vector $\mathbf{b}$, the ACME current should take the form $\mathbf{j}_{\mathrm{ACME}}=f(b^2,M,\mu)\mu_5\mathbf{b}$, where $f$ is a function depending on model parameters.
In the Weyl semimetal phase, the current for small chemical potential $\mu\ll |\mathbf{b}|-M$ and low temperature $T\ll \mu$ is~\cite{SI}
\begin{eqnarray}\label{Eq:current_Weyl_small_mu}
    \mathbf{j}_{\mathrm{ACME}}\approx \frac{M^2\mu^3}{3\pi^2\mathbf{b}^2(\mathbf{b}^2-M^2)}\mu_5\mathbf{k}_W.
\end{eqnarray}
In the limit of  large chemical potential, $\mu\gg M+|\mathbf{b}|$, we find~\cite{SI}
\begin{eqnarray}\label{Eq:current_large_mu}
    \mathbf{j}_{\mathrm{ACME}}\approx \frac{M}{2\pi}\mu_5 \mathbf{b}.
\end{eqnarray}
Equations~\eqref{Eq:current_Weyl_small_mu} and \eqref{Eq:current_large_mu} clearly show that the ACME is non-topological. 
In both limits, the current increases with the chiral symmetry breaking parameter, $M$.

\begin{figure}[t]
	\includegraphics[width=0.4\textwidth]{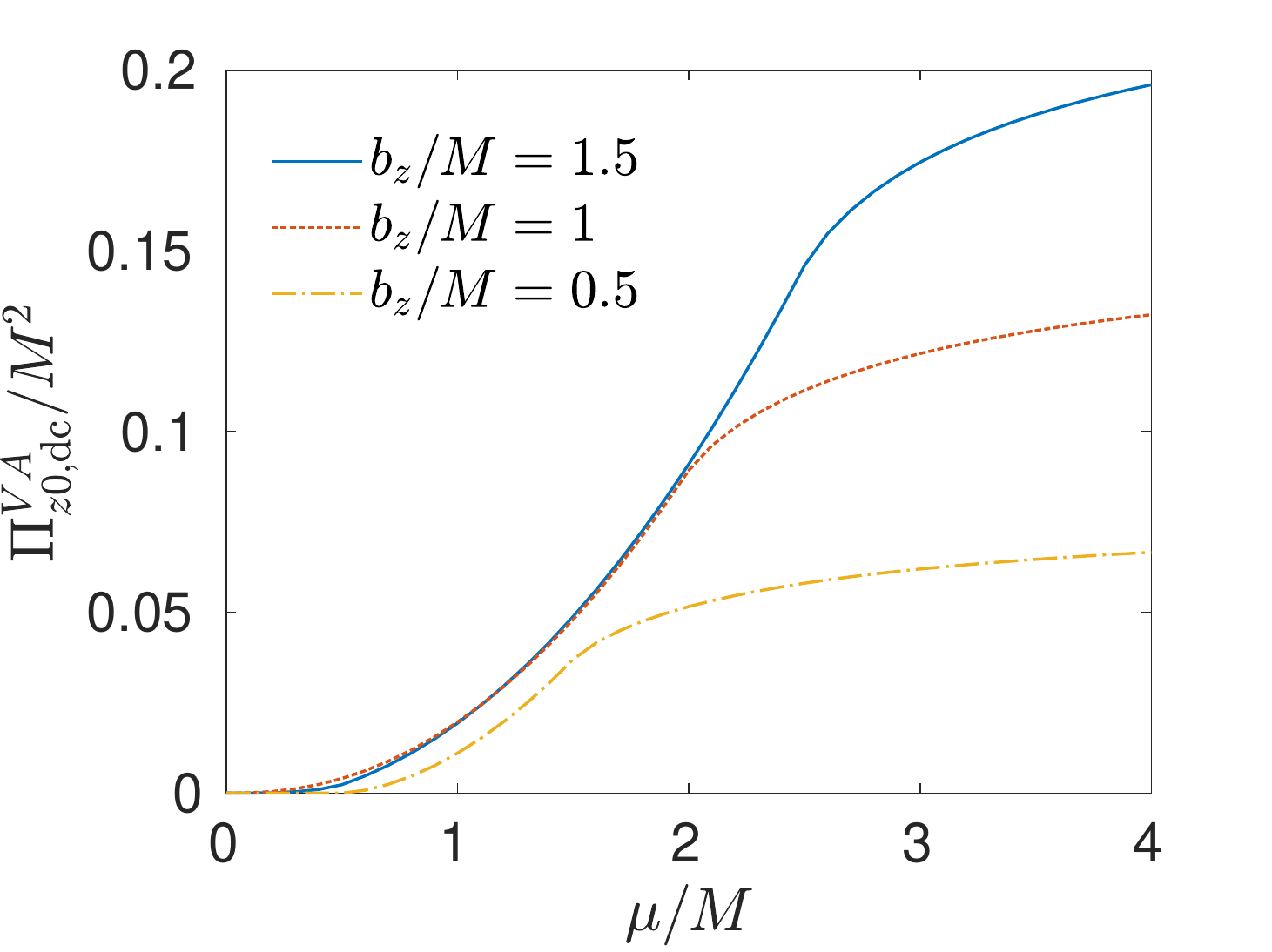}
	\caption{The dc response function as a function of $\mu/M$ for different $b_z$ corresponding to the Weyl semimetal ($b_z/M=1.5$) and gapped ($b_z/M=0.5$) phases as well as the phase boundary ($b_z/M=1$).}\label{Fig:Pii0_omega0}
\end{figure}

Figure~\ref{Fig:Pii0_omega0} shows the dc response function for different parameters as a function of the chemical potential. 
As in Fig.~\ref{Fig:Pii0_omega}, $\mathbf{b}$ is taken to be along the $z$-direction. 
For large $\mu$, the response function scales as $b_z$ in both phases, as predicted by  Eq.~\eqref{Eq:current_large_mu}. 

The dc response function is obtained by first taking the $\mathbf{q}=0$ limit and then the $\omega=0$ limit. 
The reverse procedure, i.e., taking $\omega\to 0$ before $\mathbf{q}\to 0$, gives thermodynamic quantities~\cite{PhysRevB.88.125105,PhysRevB.91.115203} and the result should be zero since the equilibrium current vanishes. 
Indeed, we find that in this case, both interband and intraband processes contribute to the response function, and they cancel with each other exactly~\cite{SI}. 

\emph{Experimental implication.\textemdash}Here we discuss how our prediction could be detected in $\mathcal{T}$-broken Weyl semimetals, see Fig.~\ref{Fig:setup} for the proposed experimental setup. 
The Weyl nodes separation is assumed to be along the $z$-direction, and then Eq.~\eqref{Eq:current_Weyl_small_mu} can be written as~\cite{SI} $j_{{\mathrm{ACME}},z}=M^2n\mu_5/b^3$, with $n$ being the particle density. 
The parallel electric and magnetic fields induce a chiral imbalance through the chiral anomaly. 
The chiral density in the steady state is
$n_{5}=\tau_{v}\mathbf{E}\cdot\mathbf{B}/(2\pi^2)$, where $\tau_{v}$ is the inter-node scattering time~\cite{PhysRevB.88.104412,PhysRevB.93.075114}. 
In the experimentally realistic limit~\cite{SI}, the chiral density $n_5$ is much smaller than the particle density $n$, such that the chiral chemical potential is proportional to the chemical potential~\cite{PhysRevD.78.074033}, $\mu_5\approx \mu n_5/(3n)$. 
Therefore the  ACME current takes the form,
\begin{eqnarray}\label{Eq:ACME_current}
    j_{\mathrm{ACME},z}=\chi\mathbf{B}\cdot\mathbf{E},
\end{eqnarray}
where $\chi=\mu M^2 e^3\tau_{v}/(6\pi^2\hbar^5 b^3v^2_F)$ and the electric charge $e$,  Fermi velocity $v_F$, and $\hbar$ are recovered.

If $\mathbf{E}$ and $\mathbf{B}$ are along the $z$-direction, the ACME current, Eq.~\eqref{Eq:ACME_current}, leads to a LMR for weak magnetic field~\cite{SI},
\begin{eqnarray}
    \frac{\delta \rho_{zz}}{\rho_{zz}}\approx-\frac{\chi B_z}{\sigma_{zz}},
\end{eqnarray}
where $\delta\rho_{zz}=\rho_{zz}(B)- \rho_{zz}$ is the resistivity change caused by the magnetic field, and $\sigma_{zz}$ is the conductivity in the zero field limit. 
The ACME provides a new microscopic  mechanism for antisymmetric LMR~\cite{PhysRevB.94.241105,PhysRevB.95.245128,PhysRevB.96.045112,PhysRevB.101.201410,zyuzin2021linear}, which was observed very recently in magnetic materials~\cite{Wang_2020} including Weyl semimetal~\cite{PhysRevLett.126.236601}. 
Taking into account the CME enhanced conductivity, $\sigma_{\mathrm{CME}}=\alpha B^2_z$ with $\alpha=e^4\mu\tau_v/(12\pi^4\hbar^4 n)$~\cite{PhysRevB.88.104412}, we find a  minimal of $\sigma_{zz}$ at the critical magnetic field strength $B_{c,z}=-\chi/(2\alpha)$.

If $\mathbf{E}$ and $\mathbf{B}$ are not parallel to the Weyl nodes separation, Eq.~\eqref{Eq:ACME_current} indicates the planar Hall conductivities~\cite{SI}
\begin{eqnarray}
    \sigma^{\mathrm{PH}}_{zx}=\chi B_x,~~ \sigma^{\mathrm{PH}}_{zy}=\chi B_y.
\end{eqnarray}
Note that the planar Hall conductivity is nonzero even if the electric and magnetic fields are collinear~\cite{PhysRevB.99.115121}.
This is different from the anomaly induced planar Hall effect~\cite{PhysRevB.96.041110,PhysRevLett.119.176804,PhysRevB.97.201110,PhysRevB.98.161110}, which requires non-parallel $\mathbf{E}$ and $\mathbf{B}$.

Now we estimate the order of magnitude of the effect. 
We take $b=0.4\si{\AA^{-1}}$, $v_F= 10^6\si{m/s}$, and $M^2/(\hbar v_F b)^2=0.75$. 
The Weyl nodes are located at $\pm k_{W,z}=\pm0.2$\si{\AA^{-1}}, and the Fermi velocities are $v_x=v_y=v_F$ and $v_z=v_F/2$. 
Assuming $\mu=0.1\si{eV}$ and $\tau_v=1\si{ns}$~\cite{PhysRevB.93.075114}, we find that the planar Hall conductivity is about $2$\si{m\Omega^{-1}cm^{-1}} for  $B_x=1$\si{T} and the critical magnetic field is $B_{c,z}\approx -0.3$\si{T}.

\emph{Conclusion.\textemdash}We proposed the ACME as a mechanism to generate dissipationless electric current in time reversal symmetry breaking Weyl semimetals. 
The current flows along the  Weyl nodes separation and is proportional to the chiral chemical potential. 
In contrast to the CME, the ACME is a Fermi surface property and is not related to the chiral anomaly. 
We show that in the presence of parallel electric and magnetic fields, the ACME results in unusual linear magnetoresistance and planar Hall conductivity, which can be measured experimentally.
The existence of the ACME current is rooted in the symmetry of the system, we thus expect the mechanism is general and can be realized in other time reversal symmetry breaking materials.

\emph{Acknowledgments.\textemdash}L.L. is grateful to P. O. Sukhachov and A. V. Balatsky for discussions and comments. This work was supported by Nordita.

\bibliography{ms.bib}
\end{document}

\title{Supplemental Material to ``Anomalous chiral magnetic effect in time reversal symmetry breaking Weyl semimetals"}

%\title{Supplemental Material to ``Anomalous chiral magnetic effect"}

%\title{Supplemental Material to ``Anomalous chiral magnetic effect induced linear magnetoresistance and planar Hall effect"}

\author{Long Liang}
\affiliation{Nordita, KTH Royal Institute of Technology and Stockholm University, Roslagstullsbacken 23, SE-106 91 Stockholm, Sweden}

\maketitle
\section{Green's function}

It is useful to find the symmetries of the Hamiltonian,
\begin{equation}\label{Eq:Hamiltonian}
    H(\mathbf{k})=\sum_{i}(k_i+b_i\gamma_5)\gamma^0\gamma^i+M\gamma^0.
\end{equation}
A Hamiltonian is inversion ($\mathcal{P}$), time reversal ($\mathcal{T}$), and particle-hole ($\mathcal{C}$) invariant if there exist unitary transformations $U_{\mathcal{P}}$,  $U_{\mathcal{T}}$, and $U_{\mathcal{C}}$ such that
\begin{eqnarray}
    U^\dag_{\mathcal{P}} H(\mathbf{k})U_{\mathcal{P}}&=&H(-\mathbf{k}),\\
    U^\dag_{\mathcal{T}} H^\ast(\mathbf{k})U_{\mathcal{T}}&=&H(-\mathbf{k}),\label{Eq:U_T}\\
    U^\dag_{\mathcal{C}} H^\ast(\mathbf{k})U_{\mathcal{C}}&=&-H(-\mathbf{k}).
\end{eqnarray}
The above relations are easily derived using the second quantized formalism~\cite{Ryu_2010}. 
It is straightforward to check that the Hamiltonian, Eq.~\eqref{Eq:Hamiltonian},
has both $\mathcal{P}$ and $\mathcal{C}$ symmetries.  
The $\mathcal{P}$ symmetry can be represented by  $U_\mathcal{P}=\gamma^0$.
We can use representations of gamma matrices such that $\gamma^2$ is pure imaginary  and other gamma matrices are real, then $U_\mathcal{C}$ can be chosen as $U_\mathcal{C}=i\gamma^2$. It is impossible to find a unitary matrix $U_\mathcal{T}$ satisfying Eq.~\eqref{Eq:U_T} when the Zeem-like field $\mathbf{b}$ is nonvanishing. For $\mathbf{b}=0$, $U_{\mathcal{T}}$ can be chosen as $i\gamma^1\gamma^3$. 

The chemical potential term is represented by $\mu I$ with $I$ being the identity matrix, so  a nonzero chemical potential breaks the $\mathcal{C}$ symmetry. The current operator $\hat{\mathbf{j}}(\mathbf{k})=\partial_{\mathbf{k}}H(\mathbf{k})$ satisfies $U^\dag_{\mathcal{C}}\hat{\mathbf{j}}^\ast(\mathbf{k})U_{\mathcal{C}}=\hat{\mathbf{j}}(-\mathbf{k})$, which means the current changes sign after $\mathcal{C}$ transformation; $U^\dag_{\mathcal{C}}\gamma_5^\ast U_{\mathcal{C}}=-\gamma_5$ such that chiral density is invariant under $\mathcal{C}$. Under the $\mathcal{P}$ transformation, the perturbative Hamiltonian,
\begin{eqnarray}
    H'=\mu_5(\mathbf{r},t)\gamma_5,
\end{eqnarray}
changes to $U^\dag_{\mathcal{P}}H'U_{\mathcal{P}}=-\mu_5(-\mathbf{r},t)\gamma_5$. 
Therefore if the chiral chemical potential is not an odd function of the position, this perturbation term breaks $\mathcal{P}$. 

Combining $\mathcal{P}$ and $\mathcal{C}$, we have
\begin{eqnarray}
    U^\dag_\mathcal{P}U^\dag_\mathcal{C}H^\ast(\mathbf{k})U_{\mathcal{C}}U_\mathcal{P}=-H(\mathbf{k}).
\end{eqnarray}
Since $H^\ast(\mathbf{k})$ and $H(\mathbf{k})$ have the same eigenvalues, we find that the eigenvalues of $H(\mathbf{k})$ come in pairs $\pm\epsilon_{s,\mathbf{k}}$, and $\epsilon_{s,\mathbf{k}}$ can be obtained through $H^2(\mathbf{k})$,
\begin{eqnarray}
   H^2(\mathbf{k})=\mathbf{k}^2+M^2+\mathbf{b}^2+2\hat{A}_{\mathbf{k}},
\end{eqnarray}
where $\hat{A}_{\mathbf{k}}=(\mathbf{k}\cdot \mathbf{b}+Mb_i\gamma^i)\gamma_5$ and $\hat{A}^2_{\mathbf{k}}=A^2_{\mathbf{k}}I$ with $A_{\mathbf{k}}=\sqrt{(\mathbf{k}\cdot\mathbf{b})^2+M^2\mathbf{b}^2}$. 
Assuming the eigenstate corresponding to the eigenvalue $t\epsilon_{s,\mathbf{k}}$ is $|\psi_{s,t}(\mathbf{k})\rangle$, then  
the projection operator $P_{s,t}(\mathbf{k})=|\psi_{s,t}(\mathbf{k})\rangle\langle \psi_{s,t}(\mathbf{k})|$ can be constructed in the following way: 
We first apply the projection operator $(1-s \hat{A}_\mathbf{k}/A_\mathbf{k})/2$ such that a state is projected to the subspace spanned by $|\psi_{s,-}\rangle$ and $|\psi_{s,+}\rangle$.
To pick up a state with definite $t$-index, we then apply the projection operator $(1-t H/\epsilon_s)/2$ to the subspace. In this way, the projection operator can be written as
\begin{eqnarray}
     P_{s,t}(\mathbf{k})=\frac{1}{4}\bigg(1-t\frac{H(\mathbf{k})}{\epsilon_{s,\mathbf{k}}}\bigg)\bigg(1-s
    \frac{\hat{A}_{\mathbf{k}}}{A_{\mathbf{k}}}\bigg).
\end{eqnarray}
Using $U_{\mathcal{C}}U_{\mathcal{P}}$, we can write the negative  energy states in terms of the positive energy ones,
\begin{eqnarray}\label{Eq:psi_ne}
   |\psi^\ast_{s,-}(\mathbf{k})\rangle=U_{\mathcal{C}}U_{\mathcal{P}}|\psi_{s,+}(\mathbf{k})\rangle=i\gamma^2\gamma^0|\psi_{s,+}(\mathbf{k})\rangle.
\end{eqnarray}
This will be useful later.

Here we give explicit expression for the Green's function,
\begin{eqnarray}
	G(k)&=&\sum_{s,t}\frac{P_{s,t}(\mathbf{k})}{i\omega_n-\mu -t\epsilon_{s,\mathbf{k}}},\\
	&=&
	\frac{1}{4}\sum_{s,t}\frac{1}{i\omega_n-\mu -t\epsilon_{s,\mathbf{k}}}\bigg[
	\gamma^\mu C_{\mu,s,t,\mathbf{k}}+m_{s,t,\mathbf{k}}+\bigg(
	\gamma^\mu C_{5,\mu,s,t,\mathbf{k}}+m_{5,s,t,\mathbf{k}}+B_{i\mu,s,t,
		\mathbf{k}}\gamma^i\gamma^\mu
	\bigg)\gamma_5
	\bigg]\gamma^0,
\end{eqnarray}
where
\begin{eqnarray}
	&&C_{\mu,s,t,\mathbf{k}}=\bigg(1, t\frac{k_i}{\epsilon_{s,{\mathbf{k}}}}+st\frac{\mathbf{k}\cdot\mathbf{b}}{\epsilon_{s,{\mathbf{k}}}A_\mathbf{k}}b_i\bigg),\\
	&&C_{5,\mu,s,t,\mathbf{k}}=\bigg(s\frac{\mathbf{k}\cdot\mathbf{b}}{A_\mathbf{k}}, t\frac{b_i}{\epsilon_{s,{\mathbf{k}}}}+st\frac{\mathbf{k}\cdot\mathbf{b}}{\epsilon_{s,{\mathbf{k}}}A_\mathbf{k}}k_i+st\frac{M^2}{\epsilon_{s,{\mathbf{k}}}A_\mathbf{k}}b_i\bigg),\\
	&&m_{s,t,\mathbf{k}}=-st\frac{M\mathbf{b}^2}{\epsilon_{s,{\mathbf{k}}}A_\mathbf{k}}-t\frac{M}{\epsilon_{s,{\mathbf{k}}}},
	~~m_{5,s,t,\mathbf{k}}=-st\frac{M\mathbf{k}\cdot\mathbf{b}}{\epsilon_{s,{\mathbf{k}}}A_\mathbf{k}},\\
	&&B_{ij,s,t,\mathbf{k}}=-st\frac{Mk_ib_j}{\epsilon_{s,{\mathbf{k}}}A_\mathbf{k}},
	~~B_{i0,s,t,\mathbf{k}}=s\frac{Mb_i}{A_\mathbf{k}}.
\end{eqnarray}
Note that $g^{i\mu}B_{i\mu,s,t}=-m_{5,s,t}$.  
In the following, the subscripts $s$, $t$ and $s'$, $t'$ will be omitted without confusion. 
The above form of the Green's function is convenient to analytically calculate the correlation function because we can use the traces of gamma matrices,
\begin{eqnarray}
\mathrm{tr} (\gamma^\mu\gamma^\nu)&=&4g^{\mu\nu},\\
\mathrm{tr}( \gamma^\mu\gamma^\nu\gamma^\rho\gamma^\sigma)&=&4(g^{\mu\nu}g^{\rho\sigma}-g^{\sigma\nu}g^{\mu\rho}+g^{\rho\nu}g^{\sigma\mu}),\\
\mathrm{tr}(\gamma_5 \gamma^\mu\gamma^
	\nu\gamma^\rho\gamma^\sigma)&=&-4i\varepsilon^{\mu\nu\rho\sigma},
\end{eqnarray}
with $g^{\mu\nu}=\mathrm{diag}(1,-1,-1,-1)$ and $\varepsilon^{0123}=1$.

\section{Anomalous chiral magnetic effect}
The anomalous chiral magnetic effect (ACME) is characterized by the  vector current--chiral density correlation function $\Pi^{VA}_{i0}$. In this section we provide detailed calculations of this function.
\begin{eqnarray}
\Pi^{VA}_{i0}(i\nu_m,\mathbf{q})&=&\frac{1}{\beta}\sum_{k}\mathrm{tr}\hat{j}_i(\mathbf{k-q}/2) G(k)\gamma_5G(k-q),\\
&=&\sum_{s,t,s',t'}\frac{f(t\epsilon_{s,{\mathbf{k}}})-f(t'\epsilon_{s',{\mathbf{k-p}}})}{i\nu_m+t'\epsilon_{s',{\mathbf{k-p}}}-t\epsilon_{s,{\mathbf{k}}}}\langle \psi_{s',t'}(\mathbf{k-q})|\hat{j}_i(\mathbf{k}-\mathbf{q}/2)|\psi_{s,t}(\mathbf{k})\rangle\langle \psi_{s,t}(\mathbf{k})|\gamma_5| \psi_{s',t'}(\mathbf{k-q})\rangle,\label{Eq:Pi_VA_transition}\\
&\equiv&
\Pi^{VA,\mathrm{even}}_{i 0}(i\nu_m,\mathbf{q})+\Pi^{VA,\mathrm{odd}}_{i 0}(i\nu_m,\mathbf{q}).
%&=&
%\frac{1}{16}\sum_{s,t,s',t'}\frac{f(t\epsilon_{s,{\mathbf{k}}})-f(t'\epsilon_{s',{\mathbf{k-p}}})}{ip_0+t'\epsilon_{s',{\mathbf{k-p}}}-t\epsilon_{s,{\mathbf{k}}}}\bigg[
%4(g^{i\rho}g^{0 l}-g^{il}g^{0 \rho})(m_{s,t,\mathbf{k}}B_{l\rho,s',t',\mathbf{k-p}}
%+
%B_{l\rho,s,t,\mathbf{k}}m_{s',t',\mathbf{k-p}})
%\nonumber\\
%&&
%-
%4(g^{i\rho}g^{0\sigma}-g^{i 0}g^{\rho\sigma}+g^{i\sigma}g^{0\rho})
%(C_{\rho,s,t,\mathbf{k}}C_{5,\sigma,s',t',\mathbf{k-p}}+ C_{5,s,t,\rho,\mathbf{k}}C_{\sigma,s',t',\mathbf{k-p}})\nonumber\\
%%
%&&+4i\varepsilon^{i\rho 0 \sigma}( C_{\rho,s,t,\mathbf{k}} C_{\sigma,s',t',\mathbf{k-p}}
%+ C_{5,\rho,s,t,\mathbf{k}}C_{5,s',t',\sigma,\mathbf{k-p}})\nonumber\\
%&&
%+4i\varepsilon^{i 0 l\rho} (m_{5,s,t,\mathbf{k}}B_{l\rho,s',t',\mathbf{k-p}}+m_{5,s',t',\mathbf{k-p}}B_{l\rho,s,t,\mathbf{k}})-\mathrm{tr}[\gamma^i \gamma^l\gamma^\rho \gamma^0 \gamma^n\gamma^\sigma\gamma_5]B_{l\rho,s,t,\mathbf{k}}B_{n\sigma,s',t',\mathbf{k-p}}
%\bigg].
\end{eqnarray}
The parity even term is 
\begin{eqnarray}
&&\Pi^{VA,\mathrm{even}}_{i 0}(i\nu_m,\mathbf{q})\nonumber\\
&&=
\frac{1}{4}\sum_{s,t,s',t'}\frac{f(t\epsilon_{s,{\mathbf{k}}})-f(t'\epsilon_{s',{\mathbf{k-p}}})}{i\nu_m+t'\epsilon_{s',{\mathbf{k-p}}}-t\epsilon_{s,{\mathbf{k}}}}\bigg[
(C_{i,\mathbf{k}}C_{5,0,\mathbf{k-p}}+ C_{5,i,\mathbf{k}}C_{0,\mathbf{k-p}}
+(i\leftrightarrow 0)
)
%\nonumber\\
%
%&&
+m_\mathbf{k}B_{i0,\mathbf{k-p}}
+
B_{i0,\mathbf{k}}m_\mathbf{k-p}
\bigg],\\
&&=\frac{1}{2}\sum_{s,t,s',t'}\frac{(t'\epsilon_{s',{\mathbf{k-p}}}-t\epsilon_{s,{\mathbf{k}}})f(t\epsilon_{s,{\mathbf{k}}})}{\nu^2_m+(t'\epsilon_{s',{\mathbf{k-p}}}-t\epsilon_{s,{\mathbf{k}}})^2}\bigg[
(C_{i,\mathbf{k}}C_{5,0,\mathbf{k-p}}+ C_{5,i,\mathbf{k}}C_{0,\mathbf{k-p}}
+(i\leftrightarrow 0)
)
%\nonumber\\
%
%&&
+m_\mathbf{k}B_{i0,\mathbf{k-p}}
+
B_{i0,\mathbf{k}}m_\mathbf{k-p}
\bigg],
\end{eqnarray}
and the parity odd term is
\begin{eqnarray}
\Pi^{VA,\mathrm{odd}}_{i 0}(i\nu_m,\mathbf{q})%=\frac{1}{16}\sum_{s,t,s',t'}\frac{f(t\epsilon_{s,{\mathbf{k}}})-f(t'\epsilon_{s',{\mathbf{k-p}}})}{ip_0+t'\epsilon_{s',{\mathbf{k-p}}}-t\epsilon_{s,{\mathbf{k}}}}\bigg[4i\varepsilon^{i\rho 0 \sigma}( C_{\rho,\mathbf{k}} C_{\sigma,\mathbf{k-p}}+ C_{5,\rho,\mathbf{k}}C_{5,\sigma,\mathbf{k-p}})\\
%&&+4i\varepsilon^{i 0 l\rho} (m_{5,\mathbf{k}}B_{l\rho,\mathbf{k-p}}+m_{5,\mathbf{k-p}}B_{l\rho,\mathbf{k}})-\mathrm{tr}[\gamma^i \gamma^l\gamma^\rho \gamma^0 \gamma^n\gamma^\sigma\gamma_5]B_{l\rho,\mathbf{k}}B_{n\sigma,\mathbf{k-p}}\bigg],\\
&=&\frac{i}{4}\sum_{s,t,s',t'}\sum_{j,l}\frac{f(t\epsilon_{s,{\mathbf{k}}})-f(t'\epsilon_{s',{\mathbf{k-p}}})}{i\nu_m+t'\epsilon_{s',{\mathbf{k-p}}}-t\epsilon_{s,{\mathbf{k}}}}\varepsilon^{0ijl}\bigg[ C_{j,\mathbf{k}} C_{l,\mathbf{k-p}}
+ C_{5,j,\mathbf{k}}C_{5,l,\mathbf{k-p}}-B_{j0,\mathbf{k}}B_{l0,\mathbf{k-p}}\nonumber\\
&&
-(B_{ji,\mathbf{k}}-B_{ij,\mathbf{k}} )(B_{li,\mathbf{k-p}}-B_{il,\mathbf{k-p}})
\bigg],\\
&=&
\frac{1}{4}\sum_{s,t,s',t'}\sum_{j,l}\frac{\nu_m[f(t\epsilon_{s,{\mathbf{k}}})-f(t'\epsilon_{s',{\mathbf{k-p}}})]}{\nu^2_m+(t'\epsilon_{s',{\mathbf{k-p}}}-t\epsilon_{s,{\mathbf{k}}})^2}\varepsilon^{0ijl}\bigg[C_{j,\mathbf{k}} C_{l,\mathbf{k-p}}
+ C_{5,j,\mathbf{k}}C_{5,l,\mathbf{k-p}}-B_{j0,\mathbf{k}}B_{l0,\mathbf{k-p}}\nonumber\\
&&
-(B_{ji,\mathbf{k}}-B_{ij,\mathbf{k}} )(B_{li,\mathbf{k-p}}-B_{il,\mathbf{k-p}})
\bigg].
\end{eqnarray}
As explained in the main text, the correlation function vanishes if the chemical potential is zero. This can also be checked explicitly by using the above expressions. 
The parity odd term vanishes in the zero frequency limit and is not of our interest. In the following we calculate the parity even term, and omit the superscript.

\subsection{uniform limit}
In this section we calculate the response function $\Pi^{VA}_{i 0}(i\nu_m,\mathbf{q})$ in the uniform limit, $\mathbf{q=0}$. 
Before calculating, we note that
in the $\mathbf{q=0}$ limit, the transition amplitude $\mathcal{M}=\langle \psi_{s,-}(\mathbf{k})|\gamma^5| \psi_{s,+}(\mathbf{k})\rangle$ can be written as [cf. Eqs.~\eqref{Eq:psi_ne} and \eqref{Eq:Pi_VA_transition}]
\begin{eqnarray}
   \mathcal{M}=
    \langle \psi_{s,-}(\mathbf{k})|i\gamma^5\gamma^0\gamma^2 |\psi^\ast_{s,-}(\mathbf{k})\rangle=(\langle \psi^\ast_{s,-}(\mathbf{k})|(i\gamma^5\gamma^0\gamma^2)^\dag |\psi_{s,-}(\mathbf{k})\rangle)^\ast=-\mathcal{M}.
\end{eqnarray}
Therefore in the uniform limit, transitions between two states with opposite energy are forbidden. 
Direct calculations also confirm this. 

Assuming the chemical potential is positive, we find
\begin{eqnarray}
\Pi^{VA}_{i 0}(i\nu_m)&=&
\sum_{\mathbf{k},t=\pm}\frac{(t\epsilon_{+,\mathbf{k}} - \epsilon_{-,\mathbf{k}})f(\epsilon_{-,\mathbf{k}})}{\nu^2_m+(t\epsilon_{+,\mathbf{k}} - \epsilon_{-,\mathbf{k}})^2}
\bigg[
\frac{M^2b_i}{A_\mathbf{k}}\frac{\mathbf{b}^2-A_\mathbf{k}}{\epsilon_{-,{\mathbf{k}}}A_\mathbf{k}}
+t
\frac{M^2b_i}{A_\mathbf{k}}\frac{\mathbf{b}^2+A_\mathbf{k}}{\epsilon_{+,{\mathbf{k}}}A_\mathbf{k}}
\bigg]\nonumber\\
&&
+\sum_{\mathbf{k},t=\pm}\frac{(t\epsilon_{-,\mathbf{k}} - \epsilon_{+,\mathbf{k}})f(\epsilon_{+,\mathbf{k}})}{\nu^2_m+(t\epsilon_{-,\mathbf{k}} - \epsilon_{+,\mathbf{k}})^2}
\bigg[
\frac{M^2b_i}{A_\mathbf{k}}\frac{\mathbf{b}^2+A_\mathbf{k}}{\epsilon_{+,{\mathbf{k}}}A_\mathbf{k}}
+t
\frac{M^2b_i}{A_\mathbf{k}}\frac{\mathbf{b}^2-A_\mathbf{k}}{\epsilon_{-,{\mathbf{k}}}A_\mathbf{k}}
\bigg],\\
&=&	\sum_{\mathbf{k},t=\pm}\frac{(t\epsilon_{+,\mathbf{k}} - \epsilon_{-,\mathbf{k}})[f(\epsilon_{-,\mathbf{k}})-f(\epsilon_{+,\mathbf{k}})]}{\nu^2_m+(t\epsilon_{+,\mathbf{k}} - \epsilon_{-,\mathbf{k}})^2}
\bigg[
\frac{M^2b_i}{A_\mathbf{k}}\frac{\mathbf{b}^2-A_\mathbf{k}}{\epsilon_{-,{\mathbf{k}}}A_\mathbf{k}}
+t
\frac{M^2b_i}{A_\mathbf{k}}\frac{\mathbf{b}^2+A_\mathbf{k}}{\epsilon_{+,{\mathbf{k}}}A_\mathbf{k}}
\bigg].\label{Eq:Pii0_VA_omega}
\end{eqnarray}
The real frequency response function is obtained by replacing $i\nu_m$ by $\omega+i0^+$.

In the zero frequency limit, Eq.~\eqref{Eq:Pii0_VA_omega} becomes
\begin{eqnarray}\label{Eq:Pi_dc}
\Pi^{VA}_{i0,\mathrm{dc}}
&=&	\sum_{\mathbf{k},t=\pm}\frac{f(\epsilon_{-,\mathbf{k}})-f(\epsilon_{+,\mathbf{k}})}{t\epsilon_{+,\mathbf{k}} - \epsilon_{-,\mathbf{k}}}
\bigg[
\frac{M^2b_i}{A_\mathbf{k}}\frac{\mathbf{b}^2-A_\mathbf{k}}{\epsilon_{-,{\mathbf{k}}}A_\mathbf{k}}
+t
\frac{M^2b_i}{A_\mathbf{k}}\frac{\mathbf{b}^2+A_\mathbf{k}}{\epsilon_{+,{\mathbf{k}}}A_\mathbf{k}}
\bigg],\\
&=&	-\sum_{\mathbf{k},s=\pm}f(\epsilon_{s,\mathbf{k}})\frac{sM^2\mathbf{b}^2b_i}{A^3_{\mathbf{k}}},\label{Eq:Pi_dc1}\\
&=&
	\sum_{\mathbf{k},s=\pm}f(\epsilon_{s,\mathbf{k}})\partial_{k_i}\rho_{5,s}(\mathbf{k}),\label{Eq:Pi_dc}
\end{eqnarray}
where
\begin{eqnarray}
	\rho_{5,s}(\mathbf{k})=\langle \psi_{s,t,\mathbf{k}} |\gamma_5 |\psi_{s,t,\mathbf{k}}\rangle=-s\frac{\mathbf{k}\cdot \mathbf{b}}{A_\mathbf{k}},
\end{eqnarray}
is the momentum resolved chiral density. Note that Eq.~\eqref{Eq:Pi_dc} is most easily obtained from Eq.~\eqref{Eq:Pi_VA_transition}.
Integrating by parts, $\Pi^{VA}_{i0,\mathrm{dc}}$ can be written as
\begin{eqnarray}
    \Pi^{VA}_{i0,\mathrm{dc}}=\sum_{\mathbf{k},s}s\frac{\mathbf{k}\cdot\mathbf{b}(A_{\mathbf{k}}k_i+s\mathbf{k}\cdot\mathbf{k}b_i)}{\epsilon_{s,\mathbf{k}}A^2_{\mathbf{k}}}f'(\epsilon_{s,\mathbf{k}}).\label{Eq:Pi_dc_FS}
\end{eqnarray}

To analytically calculate the response function, we first assume that $\mathbf{b}$ is along the $z$-direction, and then Eq.~\eqref{Eq:Pi_dc_FS} becomes
\begin{eqnarray}
    \Pi^{VA}_{z0,\mathrm{dc}}=\frac{\mu^2}{2\pi^2}\int^{1}_{x_{-,\mu}}\mathrm{d}x\frac{x\sqrt{(\mu x+b_z)^2-M^2}}{\mu x+b_z}-\frac{\mu^2}{2\pi^2}\int^{1}_{x_{+,\mu}}\mathrm{d}x\frac{x\sqrt{(\mu x-b_z)^2-M^2}}{\mu x-b_z},
\end{eqnarray}
where $x_{-,\mu}=\min\{1,\max\{-1,\frac{M-b_z}{\mu}\}\}$ and $x_{+,\mu}=\min\{1,\frac{M+b_z}{\mu}\}$. To get the above result we have assumed that the temperature is much lower than the chemical potential such that $f'(\epsilon_s)\approx-\delta(\epsilon_s-\mu)$.

We are mainly interested in the Weyl semimetal phase with $b_z>M$, and the most relevant parameter region is $\mu\ll b_z-M$. In this limit,
\begin{eqnarray}\label{Eq:Pi_dc_Weyl}
\Pi^{VA}_{z0,\mathrm{dc}}=\frac{\mu^2}{2\pi^2}\int^{1}_{-1}\mathrm{d}x\frac{x\sqrt{(\mu x+b_z)^2-M^2}}{\mu x+b_z}\approx \frac{M^2\mu^3}{3\pi^2b^2_z\sqrt{b^2_z-M^2}}.
\end{eqnarray}
Note that the total density in the low temperature and small chemical potential limit is
\begin{eqnarray}
n\approx \frac{\mu^3}{3\pi^2\sqrt{1-(M/b_z)^2}},
\end{eqnarray}
so Eq.~\eqref{Eq:Pi_dc_Weyl} can be written as
%Recovering the Fermi velocity $v_F$ and reduced Plank constant $\hbar$, we have
\begin{eqnarray}
\Pi^{VA}_{z0,\mathrm{dc}}\approx \frac{M^2 n}{b^3_z}.
\end{eqnarray}
Replacing $\partial_{k_i}\rho_{5,-}(\mathbf{k})$ by the value at Weyl points in Eq.~\eqref{Eq:Pi_dc}, we obtain the same result.

In the gapped phase ($b_z<M$), we assume the chemical potential is larger than $M-b_z$ and $\mu-(M-b_z)\ll M-b_z$, then
\begin{eqnarray}
\Pi^{VA}_{z0,\mathrm{dc}}=\frac{\mu^2}{2\pi^2}\int^{1}_{\frac{M-b_z}{\mu}}\mathrm{d}x\frac{x\sqrt{(\mu x+b_z)^2-M^2}}{\mu x+b_z}\approx \frac{(M-b_z)(\mu+b_z-M)^{3/2}}{2\pi^2\sqrt{M}}.
\end{eqnarray}

At the phase boundary ($b_z=M$) and for small chemical potential, we find
\begin{eqnarray}
\Pi^{VA}_{z0,\mathrm{dc}}=\frac{\mu^2}{2\pi^2}\int^{1}_{0}\mathrm{d}x\frac{x\sqrt{(\mu x+M)^2-M^2}}{\mu x+M}\approx \frac{\sqrt{2}\mu^{5/2}}{5\pi^2\sqrt{M}}.
\end{eqnarray}

We now consider the large chemical potential limit. 
In this limit, the relevant momentum is large such that corresponding energy dispersion becomes quadratic, and $\epsilon_{+}$ and $\epsilon_{-}$ bands can be viewed as degenerate. We find in this case
\begin{eqnarray}
\Pi^{VA}_{z0,\mathrm{dc}}=\frac{\mu^2}{2\pi^2}\int^{1}_{\frac{M-b_z}{\mu}}\mathrm{d}x\frac{x\sqrt{(\mu x+b_z)^2-M^2}}{\mu x+b_z}-\frac{\mu^2}{2\pi^2}\int^{1}_{\frac{M+b_z}{\mu}}\mathrm{d}x\frac{x\sqrt{(\mu x-b_z)^2-M^2}}{\mu x-b_z}\approx \frac{M b_z}{2\pi}.
\end{eqnarray}
This result applies to both the Weyl semimetal and gapped phases as well as the phase boundary.

Since $\mathbf{b}$ is the only vector after integration, $\Pi^{VA}_{i0,\mathrm{dc}}$ should take the form $\Pi^{VA}_{i0,\mathrm{dc}}\sim b_i/|\mathbf{b}|$.  It is thus easy to obtain the expression for the response function for a general vector $\mathbf{b}$. 
%This result can be applied to a topological insulator in the presence of a Zeeman field.

%\subsection{consistency check}
%Because of the current conservation, $\partial_t n+\nabla\cdot \mathbf{j}=0$, the correlation function $\Pi^{VA}_{00}$ and $\Pi^{VA}_{i0}$ satisfy $\partial_t \Pi^{VA}_{00}+\sum_i\partial_i\Pi^{VA}_{i0}=0$. We calculated the correlation function $\Pi^{VA}_{i0}(\omega,\mathbf{q})$ and find that the conservation law is preserved. 

As a self-consistency check, we calculated also $\Pi^{VA}_{00}$, and find the the current conservation, $\partial_t \Pi^{VA}_{00}+\sum_i\partial_i\Pi^{VA}_{i0}=0$, is satisfied. 

\subsection{static limit}
In this section we calculate the response function by taking $\omega\to 0$ limit before $\mathbf{q}\to 0$ limit. This is the static limit that gives thermodynamic property~\cite{PhysRevB.88.125105,PhysRevB.91.115203} and therefore the result should be zero. Indeed, we find
\begin{eqnarray}
\lim_{\mathbf{q}\to 0}\lim_{\omega\to 0}\Pi^{VA}_{i0}(\omega,\mathbf{q})
&=&
\frac{1}{2}\sum_\mathbf{k}f'(\epsilon_{-,{\mathbf{k}}})\bigg[
\frac{A_\mathbf{k}k_i-\mathbf{k}\cdot\mathbf{b}b_i}{\epsilon_{-,{\mathbf{k}}}A_\mathbf{k}}
\frac{\mathbf{k}\cdot\mathbf{b}}{A_\mathbf{k}}
-
\frac{A_\mathbf{k}b_i-\mathbf{k}\cdot\mathbf{b}k_i-M^2b_i}{\epsilon_{-,{\mathbf{k}}}A_\mathbf{k}}
+\frac{M^2b_i}{A_\mathbf{k}}\frac{\mathbf{b}^2-A_\mathbf{k}}{\epsilon_{-,{\mathbf{k}}}A_\mathbf{k}}
\bigg]\nonumber\\
&&-\frac{1}{2}\sum_\mathbf{k}f'(\epsilon_{+,{\mathbf{k}}})\bigg[
\frac{A_\mathbf{k}k_i+\mathbf{k}\cdot\mathbf{b}b_i}{\epsilon_{+,{\mathbf{k}}}A_\mathbf{k}}
\frac{\mathbf{k}\cdot\mathbf{b}}{A_\mathbf{k}}
+
\frac{A_\mathbf{k}b_i+\mathbf{k}\cdot\mathbf{b}k_i+M^2b_i}{\epsilon_{+,{\mathbf{k}}}A_\mathbf{k}}
-\frac{M^2b_i}{A_\mathbf{k}}\frac{\mathbf{b}^2+A_\mathbf{k}}{\epsilon_{+,{\mathbf{k}}}A_\mathbf{k}}
\bigg]\nonumber\\
&&-\sum_{\mathbf{k},s=\pm}f(\epsilon_{s,\mathbf{k}})\frac{sM^2\mathbf{b}^2b_i}{A^3_{\mathbf{k}}},\label{Eq:static}\\
&=&-\sum_\mathbf{k}\sum_s s\frac{\mathbf{k}\cdot\mathbf{b}(A_{\mathbf{k}}k_i+s\mathbf{k}\cdot\mathbf{k}b_i)}{\epsilon_{s,\mathbf{k}}A^2_{\mathbf{k}}}f'(\epsilon_{s,\mathbf{k}})
+f(\epsilon_{s,\mathbf{k}})\frac{sM^2\mathbf{b}^2b_i}{A^3_{\mathbf{k}}}=0.
\end{eqnarray}
The first two lines in Eq.~\eqref{Eq:static} describe the intraband processes, and they cancel exactly with the interband ones.

\subsection{lattice model}
In this section we calculate the response function by using a lattice model~\cite{PhysRevLett.111.027201,PhysRevLett.115.177202}.
The lattice model is obtained by replacing $k_i$ by 
	$h_i=-t_p\sin{k_i}$, and $M$ by 
	$M_\mathbf{k}=M-t_s\sum_i\cos{k_i}$, where $t_p$ and $t_s$ are hopping parameters. 
The energy dispersion  and Green's function take the same form as in the continuum case, and the current operator $\hat{j}_i$ is (repeated indices are not summed)
$\hat{j}_i=\partial_{k_i}h_i\gamma^i+\partial_{k_i}M_\mathbf{k}$. The calculations for the correlation function are similar as in the continuum model. Assuming $\mu>0$, we find
\begin{eqnarray}
\Pi^{VA}_{i0}(i\nu_m)
&=&	\sum_{\mathbf{k},s=\pm }\frac{(s\epsilon_{+,\mathbf{k}} - \epsilon_{-,\mathbf{k}})[f(\epsilon_{-,\mathbf{k}})-f(\epsilon_{+,\mathbf{k}})]}{\nu^2_m+(s\epsilon_{+,\mathbf{k}} - \epsilon_{-,\mathbf{k}})^2}
\bigg[
\frac{\mathbf{b}^2-A_\mathbf{k}}{\epsilon_{-,{\mathbf{k}}}}
+s\frac{\mathbf{b}^2+A_\mathbf{k}}{\epsilon_{+,{\mathbf{k}}}}
\bigg]\frac{M^2_\mathbf{k}b_i\partial_{k_i}h_i-M_\mathbf{k}\partial_{k_i}M_\mathbf{k}\mathbf{h}\cdot\mathbf{b}}{A^2_\mathbf{k}}.
\label{Eq:Pii0_VA_omega_lattice}
\end{eqnarray}
The $-M_\mathbf{k}\partial_{k_i}M_\mathbf{k}\mathbf{h}\cdot\mathbf{b}$ term in the above expression comes from the $\partial_{k_i}M_{\mathbf{k}}$ term in the current operator. 
The dc correlation function takes the same form as Eq.~\eqref{Eq:Pi_dc}, with $\rho_{5,s}(\mathbf{k})=-s\mathbf{h}\cdot\mathbf{b}/A_{\mathbf{k}}$.

\section{ACME induced linear magnetoresistance and Hall conductivity
}
In this section we discuss the experimental detection of the ACME in $\mathcal{T}$-broken Weyl semimetals. 
We assume that the Weyl nodes separation is along the $z$-direction. 
To generate a chiral imbalance, we apply parallel electric and magnetic fields to the sample.  
The chiral anomaly will induce a chiral density $n_5$~\cite{PhysRevB.88.104412},
\begin{eqnarray}
\partial_t n_5=\frac{e^2}{2\pi^2\hbar^2}\mathbf{E}\cdot\mathbf{B}-\frac{n_5}{\tau_v},
\end{eqnarray}
where $\tau_v$ is the inter-node scattering that leads to the relaxation of the chiral density. In the steady state,
\begin{eqnarray}
    n_5=\frac{e^2\tau_v\mathbf{E}\cdot\mathbf{B}}{2\pi^2\hbar^2}. 
\end{eqnarray}

The total density $n$ is $n=n_L+n_R\propto (\mu+\mu_5)^3+(\mu-\mu_5)^3$,
while the chiral density is $n_5=n_L-n_R\propto (\mu+\mu_5)^3-(\mu-\mu_5)^3$, so
in the limit of $\mu_5\ll\mu$, we have 
\begin{eqnarray}
    \mu_5\approx \frac{n_5\mu}{3 n}.
\end{eqnarray}
Therefore the ACME current can be written as
\begin{eqnarray}
j_{\mathrm{ACME},z}=\chi \mathbf{E}\cdot\mathbf{B},
\end{eqnarray}
where $\chi=\mu M^2 e^3\tau_{v}/(6\pi^2\hbar^5 b^3v^2_F)$. For later convenience, $v_F$ has been recovered. 

The total current can be separated into the Ohmic, anomalous Hall effect (AHE), ACME, and chiral magnetic effect (CME) parts,
\begin{eqnarray}
\mathbf{j}=\mathbf{j}_{\mathrm{Ohm}}+\mathbf{j}_{\mathrm{AHE}}+\mathbf{j}_{\mathrm{ACME}}+\mathbf{j}_{\mathrm{CME}}.
\end{eqnarray}
Since the magnetic and electric fields are parallel, there is no ordinary Hall current.
The Ohmic current is $\mathbf{j}=\sigma\mathbf{E}$, here we assume the conductivity tensor in the absence of magnetic field is isotropic.  
The anomalous Hall current is $\mathbf{j}_{\mathrm{AHE}}=\sigma_{\mathrm{AHE}}\hat{\mathbf{k}}_W\times \mathbf{E}$, where $\hat{\mathbf{k}}_W=\mathbf{k}_W/|\mathbf{k}_W|$ is the unit vector along the Weyl nodes separation. 
The CME current is~\cite{PhysRevB.88.104412} $\mathbf{j}_{\mathrm{CME}}=e^2\mu_5\mathbf{B}/(2\pi^2\hbar^2)=\alpha (\mathbf{E}\cdot\mathbf{B})\mathbf{B}$ with $\alpha=e^4\mu\tau_v/(12\pi^4\hbar^4 n)$. 

The conductivity tensor is thus
\begin{eqnarray}
\sigma(\mathbf{B})=\left(\begin{array}{ccc}
   \sigma+\alpha B^2_x  &  -\sigma_{\mathrm{AHE}}+\alpha B_xB_y & \alpha B_xB_z\\
   \sigma_{\mathrm{AHE}}+\alpha B_xB_y  & \sigma+\alpha B^2_y & \alpha B_yB_z\\
    \chi B_x+\alpha B_xB_z & \chi B_y+\alpha B_yB_z & \sigma+\chi B_z+\alpha B^2_z
\end{array}\right).
\end{eqnarray}
The off-diagonal conductivity is neither symmetric nor antisymmetric, which is known as the planar Hall effect.
We can write the conductivity tensor as a symmetric part plus an antisymmetric one,   
and the anti-symmetric part of the conductivity tensor describes the Hall response~\cite{RevModPhys.82.1539,tsirkin2021separation}. Therefore  the ACME gives rise to the Hall conductivity, 
\begin{eqnarray}
    \sigma^{\mathrm{H}}_{zx}=\frac{\chi B_x}{2},~~ \sigma^{\mathrm{H}}_{zy}=\frac{\chi B_y}{2}.
\end{eqnarray}
This is  unusual since the electric and magnetic fields are parallel. 
%Similar effect was predicted in tilted Weyl semimetals.
Moreover, there is  a minimal of $\sigma_{zz}$ at the critical magnetic field strength 
\begin{eqnarray}
    B_{c,z}=-\frac{\chi}{2\alpha}=-\frac{\pi^2M^2n}{\hbar b^3v^2_Fe}. 
\end{eqnarray}
Note that $B_{c,z}$ is independent of the inter-node scattering time $\tau_v$.
Assuming $\mathbf{E}$ and $\mathbf{B}$ are along the $z$-direction, the resistivity $\rho_{zz}$ is
\begin{eqnarray}
\rho_{zz}=\frac{1}{\sigma+\chi B_z+\alpha B^2_z}.
\end{eqnarray}
The resistivity $\rho_{zz}$ reaches a maximal at $B_{c,z}$, and the magnetoresistance in the small $B_z$ limit is proportional to $B_z$. 
%The antisymmetric linear magnetoresistance was predicted in tilted Weyl semimetals. 
The ACME provides a different mechanism for the antisymmetric linear magnetoreistance, which does not rely on the tilted Weyl cones~\cite{PhysRevB.95.245128,PhysRevB.96.045112,PhysRevB.101.201410,zyuzin2021linear}.
In~\cite{PhysRevB.94.241105}, the linear magnetoresistance was predicted in Weyl semimetals with band-bending term, and the current is found to be along the Weyl nodes separation. This thus can be understood through the ACME.
%As mentioned in the main text, this linear dependence is different from the observed $|B|$ behaviour of the magnetoresistance in Weyl and Dirac semimetals where the magnetic field is perpendicular to the current.

Now we estimate the magnitude of the effect.
The typical value of  Fermi velocity  is  $v_F= 10^6\si{m/s}$. We take $b=0.4\si{\AA^{-1}}$ and  $M^2/(\hbar v_F b)^2=3/4$, such that  the Weyl points  are  located at $\pm k_{W,z}=\pm 0.2$\si{\AA^{-1}}, and the Fermi velocities are $v_x=v_y=v_F$ and $v_z=v_F/2$. 
The chemical potential is taken to be $\mu=0.1\si{eV}$, which leads to a particle density $n\approx 2\times 10^{17}$\si{cm^{-3}}. 
The inter-nodes scattering time is taken to be $\tau_v=1\si{ns}$~\cite{PhysRevB.93.075114}.
Then for $B=1$\si{T} and $E=10$\si{V/m}~\cite{Li_2016,Arnold_2016}, the chiral density is $n_5\approx 1\times 10^{15}$\si{cm^{-3}}. Thus the assumption $n_5\ll n$ used previously is valid. Using these parameters, the Hall conductivity $\sigma^{\mathrm{H}}_{zx}$ is about $1$\si{m\Omega^{-1}cm^{-1}} for $B_x=1$\si{T} and  the critical magnetic field is estimated to be $B_{c,z}\approx -0.3$\si{T}.

\bibliography{supplement.bib}